\pdfoutput=1
\documentclass[doublecol]{epl2}
\usepackage{graphicx,color}
\usepackage{amsmath,amssymb}

\newcommand{\eref}[1]{Eq.~(\ref{#1})}%
\newcommand{\fref}[1]{Fig.~\ref{#1}} %
\newcommand{\Fref}[1]{Figure~\ref{#1}}%

\begin{document}

\title{Record Statistics of Continuous Time Random Walk}

\author{Sanjib Sabhapandit}
\institute{Raman Research Institute, Bangalore 560080, India}

\date{August 5, 2010}


\abstract{
The statistics of records for a time series generated by a continuous
time random walk is studied, and found to be independent of the
details of the jump length distribution, as long as the latter is
continuous and symmetric.  However, the statistics depend crucially on
the nature of the waiting time distribution.  The probability of
finding $M$ records within a given time duration $t$, for large~$t$,
has a scaling form, and the exact scaling function is obtained in
terms of the one-sided L\'evy stable distribution.  The mean of the
ages of the records, defined as $\langle t/M\rangle$, differs from
$t/\langle M\rangle$. The asymptotic behaviour of the shortest and the
longest ages of the records are also studied. }

\maketitle

People's interest in records is evident from the popularity of the
\textit{Guinness World Records} --- which itself holds a record as the
best selling copyright book.  An observation is called a (upper)
record if its value exceeds that of all previous observations.
Applications of records are found in diverse fields such as
meteorology~\cite{meteorology}, hydrology~\cite{hydrology},
economics~\cite{economics}, and sports~\cite{sports}.  Records also
play an important role in highlighting climate data to increase public
awareness of climate change and global warming --- e.g., a recent
analysis from the United States {\it National Climatic Data Center}
reveals that the global combined land and ocean average surface
temperatures for June, April to June average, and January to June
average, of 2010, are the warmest on record~\cite{NOAA}.

Frequently asked questions in the study of records include: (a) How
many records occur in a given duration?  (b) How long does a record
stay before it is broken by a new one?  In the case of a discrete time
series $\{x_0,x_1,x_2,\dotsc,x_N \}$ consisting of independent and
identically distributed (IID) random variables, the statistics
regarding the above questions are well-studied~\cite{Chandler:52,
Arnold:98, Nevzorov:01, Schmittmann:99} --- which have been useful in
understanding and analysing evolutions in complex systems, ranging
from phase slip in charge density waves, ageing in glassy systems, to
biological macroevolution~\cite{Sibani} and
adaptation~\cite{adaptation, adaptation-models}.  The record
statistics for independent but non-identically distributed entries
have been studied recently in the context of biological evolution and
warming climate~\cite{Krug:07, Krug:10, Krug:10-2}.

In spite of the continued interest in records, the theory has been
mostly restricted to the case of independent random variables ---
while a time series often contains correlated entries.  It is only
recently that Majumdar and Ziff~\cite{Majumdar:08} initiated the study
of record statistics for correlated entries --- by considering a time
series generated by a ``discrete time random walk'' (DTRW):
$x_i=x_{i-1} + \xi_i$ with IID jump lengths $\{\xi_i \}$.  A
subsequent study of record statistics of a random walk with Cauchy
distributed jumps and a drift has appeared in the context of a driven
particle in a random landscape~\cite{LeDoussal:09}.  These studies,
however, assume that the samples are collected at regular intervals,
say $\tau_0$, so that the number of entries $N$, in a given time
duration $t$, is fixed: $N=t/\tau_0$.  This is not the case in many
real situations where the events happen at irregular
intervals. Consequently for a fixed time duration $t$, the number of
entries in the time series $\{x(t_0), x(t_1), x(t_2),
\dotsb,x(t_N) \}$ is random and, therefore, the results for fixed $N$
would not hold.  Thus, the natural first step would be to study the
statistics of records in the simplest possible model of correlated
time series in continuous time. In this Letter we carry out this
important step.

The well-known model that takes into account of the random waiting
times $\tau_i=t_i-t_{i-1}$, between successive steps of a random walk,
is the so-called ``continuous time random walk'' (CTRW), that was
introduced by Montroll and Weiss~\cite{montroll:65}, and ever since
has been successfully used to describe anomalous diffusion in various
complex systems~\cite{Bouchaud:90,Metzler:00,Weiss:83,Weiss:94,
Hughes:95}, including finance and economics~\cite{Scalas:06}.  In this
Letter, we obtain exact asymptotic results for the statistics of the
number and the ages of records, of a time series generated by the CTRW
with the IID waiting times drawn from a probability density function
(PDF) $\rho(\tau)$, and IID jump lengths drawn from a continuous and
symmetric PDF $\phi(\xi)$.  The results are independent of $\phi(\xi)$
--- and the reason for this universality is the same as in
Ref.~\cite{Majumdar:08}, namely, the use of the Sparre-Andersen
theorem~\cite{SparreAndersen, Feller}.  However, with respect to the
nature of $\rho(\tau)$, the statistics of records within a given time
duration $t$, display rather rich scaling behaviour for large~$t$.
They are essentially determined by the behaviour of the Laplace
transform $\widetilde{\rho}(s)=\int_0^\infty e^{-s \tau} \rho(\tau) \,
d\tau =1-(\tau_0 s)^\alpha +\dotsb$ as $s\rightarrow 0$, where
$\alpha=1$ as long as the mean waiting time is finite, whereas the
latter becomes infinite for $0<\alpha<1$.  The asymptotic results of
Ref.~\cite{Majumdar:08} correspond to the special case $\alpha=1$ of
the more general model considered in this Letter.

We first summarise our main results.  We find that, the probability
$P(M,t)$ of finding $M$ records within a given time duration $t$, 
for large $M$ and $t$ with the scaled variable
$M(t/\tau_0)^{-\alpha/2}$ fixed, has the scaling form
\begin{equation}
P(M,t)\, \sim\, \left(t/\tau_0\right)^{-\alpha/2} g_\alpha \left(M
(t/\tau_0)^{-\alpha/2}\right). 
\label{P(M,t)}
\end{equation}
The scaling function is given by
\begin{equation}
  g_\alpha(x)=(2/\alpha) x^{-(1+2/\alpha)} L_{\alpha/2}
  (x^{-2/\alpha}), \quad 0<\alpha\le 1,
\label{scaling PDF}
\end{equation} 
where $L_\mu(x)$ is the one-sided ($\beta=+1$) L\'evy stable PDF.  The
asymptotic behaviour of the moments of the number of records is given
by
\begin{equation}
  \bigl\langle M^\nu \bigr\rangle \sim \frac{(2/\alpha)\Gamma (\nu)}{
    \Gamma(\nu\alpha/2)} \left(\frac{t}{\tau_0}\right)^{\nu\alpha/2}.
\label{moments}
\end{equation}
The mean of the ages of the records $\langle l\rangle=\langle
t/M\rangle$, grows as
\begin{equation}
\Bigl\langle \frac{t}{M}\Bigr\rangle
\sim \frac{\tau_0 (\alpha/2)}{ \Gamma\bigl(1-\frac{\alpha}{2}\bigr) }
\biggl[\ln
 \Bigl(\frac{t}{\tau_0}\Bigr)
-\Psi\Bigl(1-\frac{\alpha}{2}\Bigr)
\biggr]
\left(\frac{t}{\tau_0}\right)^{1-\alpha/2} 
\label{mean age}
\end{equation}
where $\Psi(x)=\Gamma'(x)/\Gamma(x)$ is the digamma function.  On the
other hand, $t/\langle M\rangle \sim \tau_0 \Gamma(1+\alpha/2)
(t/\tau_0)^{1-\alpha/2}$. As for the extreme ages of the records: the
mean shortest age grows as $\langle l_\text{min}\rangle
\sim\tau_0 \big[\Gamma(1-\alpha/2) \big]^{-1} (t/\tau_0)^{1-\alpha/2}$
whereas the mean longest age has a linear growth $\langle l_\text{max}
\rangle \sim C_\alpha\, t$  with the growth constant given by 
\begin{equation}
C_\alpha=\int_0^\infty dx\left[1+x^{\alpha/2} e^x \int_0^x
  y^{-\alpha/2} e^{-y}\,dy\right]^{-1}.
\label{Q}
\end{equation}

\begin{figure}
\includegraphics[width=\hsize]{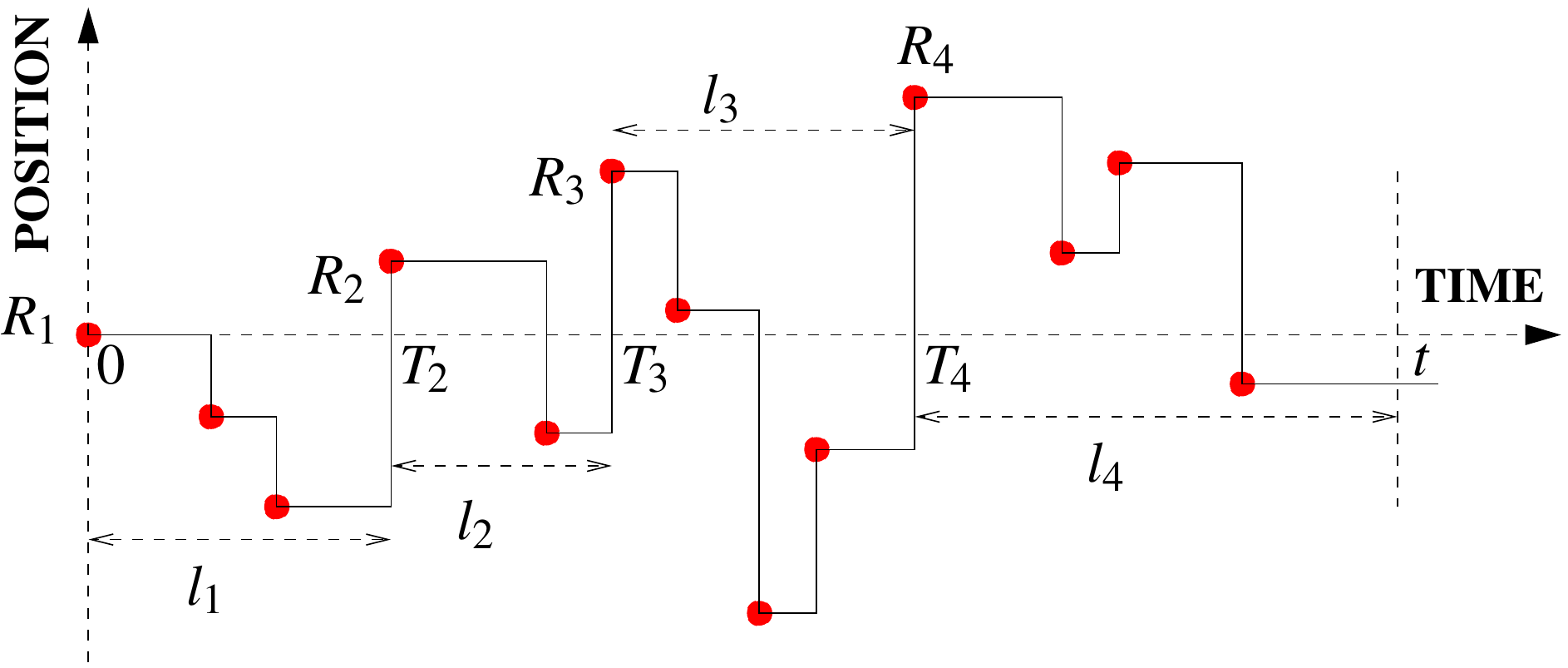}
\caption{\label{CTRW-fig} (colour on-line). A realisation of a CTRW in
  the time interval $[0,t]$. Filled circles (in red) show the
  positions of the walker immediately after the jump. The horizontal
  lines between successive steps show the waiting times, whereas the
  vertical lines show the step sizes. $R_i$ and $T_i$ denote
  respectively the value and the time of occurrence of the $i$-th
  record. $l_i$ denotes the age of the $i$-th record. The number of
  records $M=4$ for this particular realisation.}
\end{figure}

Now we proceed to obtain the joint probability distribution of the
ages and the number of records in a given duration $t$ --- from which
the statistics of individual variables can be computed by integrating
out the rest.  In this context it is useful to refer to
\fref{CTRW-fig}.  Consider a time series $\{x(0), x(t_1), x(t_2),
\dotsb,x(t_N) \}$ generated by a CTRW with the IID jump sizes
$\xi_i=x(t_i)-x(t_{i-1})$ drawn from a continuous and symmetric PDF
$\phi(\xi)$, and the IID waiting times $\tau_i=t_i-t_{i-1}$ between
successive jumps drawn from a one-sided PDF $\rho(\tau)$. We have set
the initial time $t_0=0$, without loss of generality.  Clearly, the
total number of steps $N$, taken within a fixed time duration $t$ ---
such that $t_N<t <t_{N+1}$ --- is a random variable that varies from
one realisation to another.  Now for any given realisation of the
walk, an entry $x(t_i)$ is a record iff $x(t_i) > x(t_j)$ for all
$j<i$.  Let $R_n$ and $T_n$ denote the value and the time of
occurrence, respectively, of the $n$-th record. By convention, the
first entry is taken to be the first record, i.e., $R_1=x(0)$, and
$T_1=0$.  Evidently, $R_{n+1} > R_{n}$ and $T_{n+1} > T_{n}$ for all
$n\ge 1$. Moreover, during the time interval between any two
successive records, say the $n$-th and the $(n+1)$-th, the random walk
does not exceed the earlier record value $R_n$.  Therefore, the
$(n+1)$-th record is the event in which a random walk starting from
the position $R_n$ at time $T_n$, exceeds its starting position $R_n$
at time $T_{n+1}$ for the first time.  It is evident that the PDF of
the above first-passage process --- let it be denoted by $f(l)$ ---
depends neither on the actual value $R_n$ nor on the individual times
$T_n$ and $T_{n+1}$, but solely on the time difference
$l=T_{n+1}-T_n$.  Let $M$ denote the total number of records in the
interval $[0,t]$. Then, after the time $T_M$, the random walk does not
exceed the value $R_M$ up to $t$, since there are no records in the
interval $(T_M,t]$. The probability of not exceeding the starting
position $R_M$ solely depends on the time duration $l=t-T_M$.  Let
$q(l)$ denote this survival probability.  Let $l_n$ be the age of the
$n$-th record, which is defined as: $l_n=T_{n+1}-T_n$ for $1\le n<M$,
and $l_M=t-T_M$.  Now, due to the Markov nature of the random walk,
different inter-record intervals are uncorrelated.  Thus, the joint
distribution of the ages $\{l_i\}\equiv \{l_1,l_2,\dotsc,l_M\}$ and
the number $M$ of records can be written as \begin{equation}
P(\{l_i\},M,t) = \left[\prod_{i=1}^{M-1} f(l_i)\right]\,
q(l_M) \, \delta\left(t-\sum_{i=1}^M
l_i\right), \label{joint-distribution} \end{equation} where the
$\delta$-function ensures that the ages add up to the total time
duration $t$.

The first-passage properties of the CTRW are known to be related to
that of the DTRW~\cite{montroll:65}.  Let $f_n$ be the probability
that a random walk exceeds its starting position for the first time at
the $n$-th step. Then $f(l)=\sum_{n=1}^\infty f_n\, p_n(l) $, where
$p_n(l)=\int_0^\infty d\tau_1 \dotsb \int_0^\infty d\tau_n\,
\bigl[\prod_{i=1}^n \rho(\tau_i)\bigr]\delta(l-\sum_{i=1}^n \tau_i)$,
is the PDF for the occurrence of the $n$-th step at time $l$.  It is
convenient to take a Laplace transform, which yields
$\widetilde{f}(s)=\int_0^\infty e^{-sl} f(l) \, dl=\sum_{n=1}^\infty
f_n \,[\widetilde{\rho}(s)]^n$.  Similarly $q(l)$ can be related to
the probability $q_n$ that the random walk does not exceed its
starting position up to the step $n$.  Unlike the first-passage event,
in this case the $n$-th step could occur at any time $\tau$ in the
interval $(0,l)$, and then the walker does not move for the remaining
duration $\tau_{n+1}=l-\tau$ (cf.~\fref{CTRW-fig}). The probability
that the walker remains fixed at least for a duration of $\tau_{n+1}$
is equal to $\int_{\tau_{n+1}}^\infty \rho(\tau) \, d\tau$ --- which
has the Laplace transform $s^{-1}[1-\widetilde{\rho}(s)]$.  Therefore,
the Laplace transform of $q(l)$ reads: $\widetilde{q}(s)=
s^{-1}[1-\widetilde{\rho}(s)]\sum_{n=0}^\infty q_n
[\widetilde{\rho}(s)]^n$.  Now, as long as the PDF $\phi(\xi)$ of the
jump lengths is continuous and symmetric, according to the
Sparre-Andersen theorem~\cite{SparreAndersen, Feller}, both $f_n$ and
$q_n$ are independent of $\phi(\xi)$ --- the respective generating
functions read: $\sum_{n=1}^\infty f_n z^n = 1-\sqrt{1-z}$ and
$\sum_{n=0}^\infty q_n z^n = 1/\sqrt{1-z}$.  These yield the Laplace
transforms:
\begin{equation}
\widetilde{f}(s)=1-\sqrt{1-\widetilde{\rho}(s)} 
\quad\text{and}\quad
\widetilde{q}(s)=s^{-1}\Bigl[1-\widetilde{f}(s)\Bigr].
\label{LT-fq}
\end{equation}
It is now evident that the joint distribution given by
\eref{joint-distribution},  depends only on the waiting time
distribution $\rho(\tau)$, and not the jump distribution $\phi(\xi)$.

Let us now compute the probability distribution of the number of
records by integrating out the ages in \eref{joint-distribution},
i.e., $P(M,t)=\int_0^\infty dl_1\dotsi\int_0^\infty dl_M \;
P(\{l_i\},M,t)$. In this regard, it is convenient take a Laplace
transform with respect to $t$, that disentangles the global constraint
imposed by the $\delta$-function. This yields
\begin{equation}
\int_0^\infty P(M,t)\, e^{-s t}\, dt
=\frac{1-\widetilde{f}(s)}{s} \Bigl[\widetilde{f}(s) \Bigr]^{M-1},
\label{LT1}
\end{equation}
for $M=1,2,\dotsc,\infty$.  The normalisation $\sum_M P(M,t) = 1$ can
be readily verified by summing the above equation over $M$.

The behaviour of $P(M,t)$ for large~$t$, can be extracted from the
small~$s$ behaviour of \eref{LT1} --- which, according
to \eref{LT-fq}, points to the small~$s$ behaviour of the Laplace
transform $\widetilde{\rho}(s)$.  Now, as long as the mean waiting
time is finite --- i.e., $\rho(\tau)$ decays faster than the power-law
tail $\tau^{-2}$ at large $\tau$, --- the Laplace transform behaves as
$\widetilde{\rho}(s) =1-\tau_0 s +\dotsb$ as $s\rightarrow 0$.  On the
other hand, if waiting time distribution has a slower decay,
$\rho(\tau)\sim \tau^{-(1+\alpha)}$ at large $\tau$, with
$0<\alpha<1$, the mean waiting time is infinite, and the Laplace
transform $\widetilde{\rho}(s)=1-(\tau_0 s)^\alpha +\dotsb$ as
$s\rightarrow 0$.  Considering both the above cases together,
from \eref{LT-fq} we get, $\widetilde{f}(s)\approx 1-(\tau_0
s)^{\alpha/2}$ as $s\rightarrow 0$, where $0<\alpha \le 1$.  Thus,
taking the limit of small~$s$ and large $M$ with keeping
$Ms^{\alpha/2}$ fixed in \eref{LT1}, results in
\begin{math}
\int_0^\infty P(M,t)\, e^{-s t}\, dt \approx \tau_0 (\tau_0
s)^{\alpha/2-1} e^{-M(\tau_0 s)^{\alpha/2}}.
\end{math}
This equation suggests that in the scaling limit $M\rightarrow
\infty$, $t\rightarrow \infty$ while the scaled variable
$M(t/\tau_0)^{-\alpha/2}$ is kept fixed, $P(M,t)$ should have the
scaling form given by \eref{P(M,t)}, so that
\begin{equation}
\int_0^\infty e^{-s y}\Bigl[y^{-\alpha/2}
  g_\alpha\bigl(y^{-\alpha/2}\bigr)\Bigr]\, 
 dy =
s^{\alpha/2-1} e^{-s^{\alpha/2}}.
\label{LT2}
\end{equation}
To evaluate the scaling function $g_\alpha(x)$, it is useful to note
that $e^{-s^\mu}$ is the Laplace transform of the one-sided L\'evy
stable PDF $L_\mu(y)$ ~\cite{Bouchaud:90}.  From this, it immediately
follows by differentiation that $\int_0^\infty e^{-sy} \left[y\,
L_\mu(y) \right] \, dy = \mu s^{\mu-1} e^{-s^\mu}$.  Comparing this
equation with \eref{LT2}, the scaling function can be expressed in
terms of the one-sided L\'evy stable PDF as in \eref{scaling PDF}.  The
normalisation of the PDF of the scaled number of records
$x=M(t/\tau_0)^{-\alpha/2}$ can be checked by integrating
\eref{scaling PDF} with the change of variable $y=x^{-2/\alpha}$ and
subsequently noting that $L_\mu(y)$ is normalised to unity, i.e.,
$\int_0^\infty g_\alpha(x)\, dx =\int_0^\infty L_{\alpha/2}(y)\,
dy=1$.  Incidentally, the PDF of the maximum displacement of a CTRW in
a time interval $t$, also has a scaling form similar to \eref{P(M,t)},
with the scaling function being identical to \eref{scaling
PDF} \emph{provided that the jumps are distributed according to a
narrow distribution having a finite variance}~\cite{Schehr:10}.
However, it is important to realise that the record statistics of the
CTRW is completely independent of the jump distribution as long as the
PDF $\phi(\xi)$ of the jump lengths is continuous and symmetric
---which also includes L\'evy flights where $\phi(\xi) \sim
|\xi|^{-1-\mu}$ is power-law distributed for large $|\xi|$ with $0
< \mu \le 2 $ and thus has a divergent second moment.

In general, $L_\mu(y)$ in \eref{scaling PDF} does not have a
closed-form expression.  However, expressing the right hand side of
\eref{LT2} as a series in $s$ and then evaluating the inverse Laplace
transform of the series, term by term, one gets \begin{equation}
 g_\alpha(x)=\frac{1}{\pi}\sum_{k=1}^\infty \frac{(-x)^{k-1}}{(k-1)!}\, \Gamma\Bigl(k\frac{\alpha}{2}\Bigr)\, \sin \Bigl(k\frac{\alpha}{2}\pi\Bigr).  \end{equation}
 This series representation is particularly useful for the numerical
 evaluation of $g_\alpha(x)$.  Using the behaviour of $L_\mu(y)$ for
 small $y$~\cite{Bouchaud:90}, we find that for large $x$,
\begin{equation}
g_\alpha(x) \approx\frac{(\frac{\alpha}{2} x)^{-\frac{(1-\alpha)}{(2-\alpha)}}}{\sqrt{(2-\alpha)\pi} }
\exp\left[-\left(\frac{2}{\alpha}-1\right) 
\left(\frac{\alpha}{2} x\right)^{\frac{2}{(2-\alpha)}} \right].
\end{equation}

\begin{figure}
\includegraphics[width=\hsize]{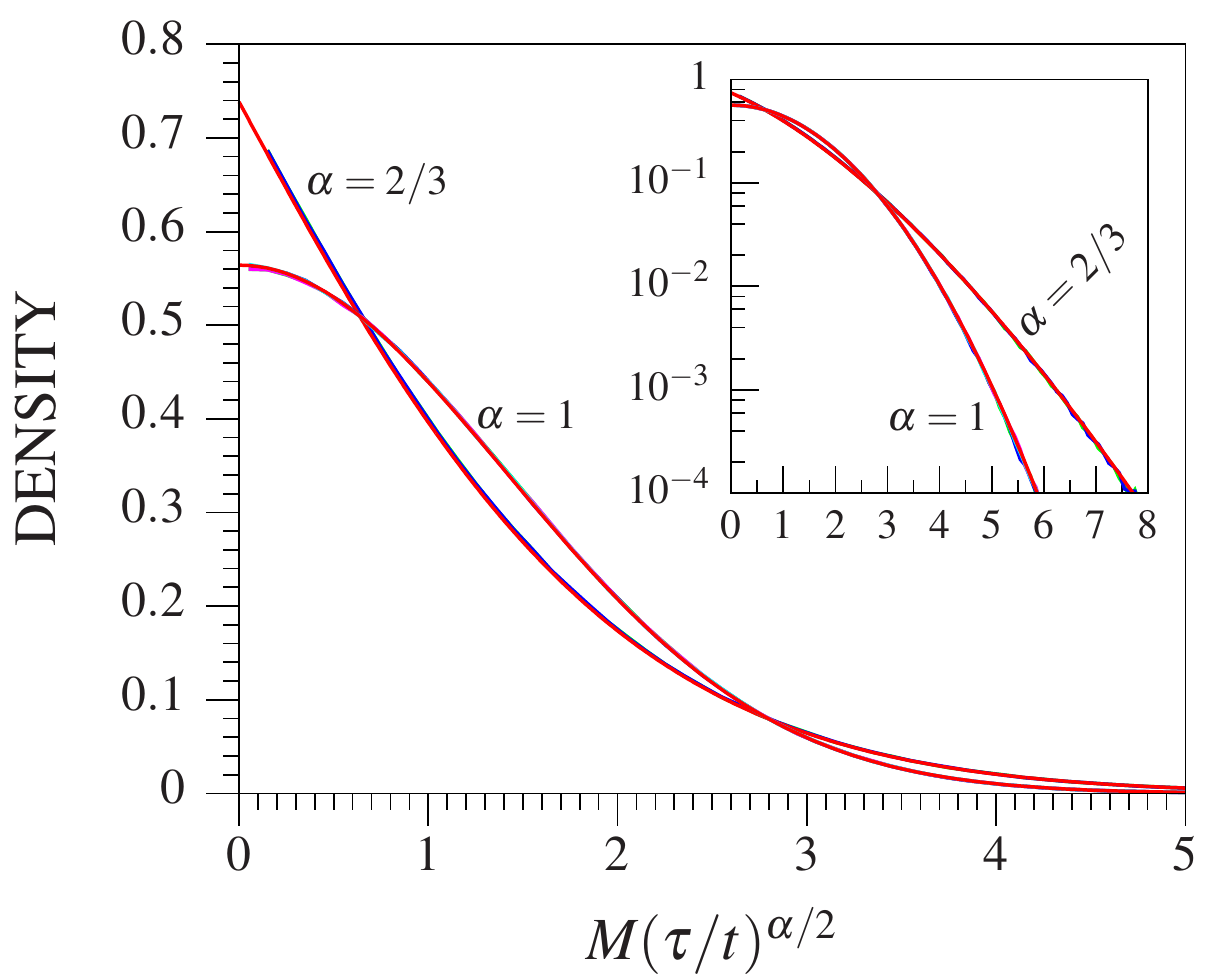}
\caption{\label{density-fig} (colour on-line). Density of the scaled
  records.  For $\alpha=2/3$, there are three curves that lie on top
  of each other: two of them (in green and blue) are obtained from
  numerical simulations (see text) and the third one (in red) plots
  the theoretical $g_{2/3}(x)$.  For $\alpha=1$ there are five curves:
  four (in green, blue, magenta and cyan) from the simulation (see
  text) and one (in red) plots $g_1(x)$.  Inset displays the curves in
  semi-log scale.}
\end{figure}

\Fref{density-fig} compares the analytical $g_\alpha(x)$ with the
numerically obtained densities of the scaled variable
$M(t/\tau_0)^{-\alpha/2}$ for $\alpha=2/3$ and $\alpha=1$.  In fact,
for these particular values of $\alpha$, one has the explicit forms:
\begin{math}
g_{2/3}(x)=(\sqrt{x}/\pi) K_{1/3} \bigl(2 (x/3)^{3/2} \bigr)
\end{math}
where $K_\nu(z)$ is the modified Bessel function, and
$g_1(x)=e^{-x^2/4}/\sqrt{\pi}$ respectively.  The latter agrees with
the scaling function found in Ref.~\cite{Majumdar:08} for the DTRW, as
expected.  Numerical simulations of the CTRW are performed with both
Gaussian and Cauchy distributed jumps to show the universality with
respect to the jump distributions.  For $\alpha=2/3$, the waiting
times are drawn from a power-law tail $\rho(\tau)\sim \tau^{-5/3}$.
For $\alpha=1$, simulations are performed with both power-law
$\rho(\tau)\sim \tau^{-5/2}$ and exponential $\rho(\tau)=e^{-\tau}$
distributions of the waiting times to show the universality of the
results for finite mean waiting time.  Each simulation is performed
with $10^7$ realisations of the CTRW, while taking $(t/\tau_0)=10^7$
for $\alpha=2/3$ and $(t/\tau_0)=10^5$ for $\alpha=1$.

To derive the moments $\langle M^\nu \rangle = \sum_M M^\nu P(M,t)$
for large~$t$, we use the scaling form given by \eref{P(M,t)}, and
subsequently replace the sum over $M$ by an integral over the scaled
variable, which is justified for large~$t$.  This gives $\langle M^\nu
\rangle \approx A_{\alpha,\nu}\, (t/\tau_0)^{\nu \alpha/2}$ with
$A_{\alpha,\nu}=\int_0^\infty x^\nu g_\alpha(x)\, dx$. Now
substituting $g_\alpha(x)$ from \eref{scaling PDF} and then making a
change of variable $y=x^{-2/\alpha}$ we get
$A_{\alpha,\nu}=\int_0^\infty y^{-\nu\alpha/2} L_{\alpha/2}(y)\, dy=
(2/\alpha)\Gamma (\nu) \bigl[\Gamma(\nu\alpha/2)\bigr]^{-1} $, where
in the last part we have used the known expression~\cite{Bouchaud:90}
for the negative moments of $L_\mu(y)$.

Let us now turn our attention to the statistics related to the ages of
the records.  The mean of the ages of the records for a given time
interval $t$, is $\langle l \rangle =\langle M^{-1} \sum_{i=1}^M
l_i\rangle= \langle t/M \rangle$.  To compute this average, it is
convenient to consider the moments generating function of
$M$. Multiplying both sides of
\eref{LT1} by $z^{M-1}$ and summing over $M$ gives $\int_0^\infty dt\,
e^{-st} \sum_{m=1}^\infty z^{M-1}
P(M,t)=s^{-1}\bigl[1-\widetilde{f}(s)\bigr]\,\bigl[1-z\widetilde{f}(s)
  \bigr]^{-1}$. Now integrating over $z$ in $[0,1]$, and
differentiating with respect to $s$, one gets 
\begin{equation}
\int_0^\infty dt\,
e^{-st} \,
\Bigl\langle \frac{t}{M}\Bigr\rangle
=\frac{\partial}{\partial s}
\left\{\frac{1-\widetilde{f}(s)}{s \widetilde{f}(s)}\;
\ln\Bigl[1-\widetilde{f}(s)
  \Bigr]\right\}.
\label{LT-mean_age}
\end{equation}
The asymptotic behaviour of the mean of the ages, for large~$t$,
emerges from the singularities near $s=0$ in the above equation.  The
inverse Laplace transform of \eref{LT-mean_age}, with
$\bigl[1-\widetilde{f}(s)\bigr]\rightarrow (\tau_0 s)^{\alpha/2}$ and
$\widetilde{f}(s)\rightarrow 1$, yields the result displayed by
\eref{mean age}. It is interesting to note that $\langle t/M \rangle$
given by \eref{mean age} differs from $t/\langle M \rangle\sim \tau_0
\Gamma(1+\alpha/2) (t/\tau_0)^{1-\alpha/2}$, where in the latter we
have used first moment from \eref{moments}.  This non-self-averaging
behaviour can be traced back to the broad distribution $f(l)\sim
l^{-(1+\alpha/2)}$ of the inter-record intervals.  It should be
emphasised that the mean of the ages of the records in a given time
interval $t$, is to be distinguished from the mean inter-record
interval --- in fact, the latter is infinite. The distinction between
the two means arises from the definition of the age of the last record
$l_M$ in a given time interval $t$ (see \fref{CTRW-fig}).

\begin{figure}[t!]
\includegraphics[width=\hsize]{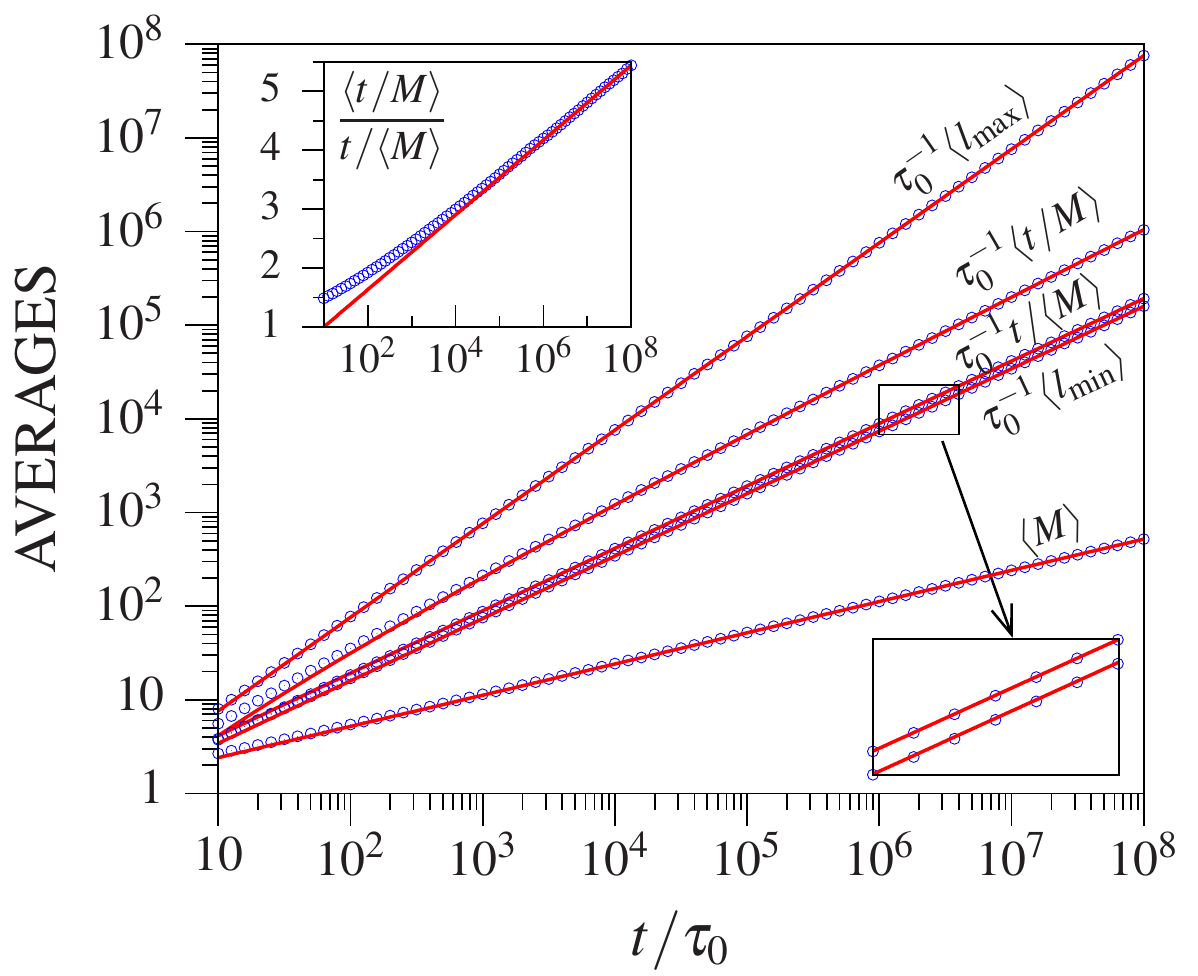}
\caption{\label{average-fig} (colour on-line).  Different averages that
  are indicated on the curves, for $\alpha=2/3$.  The points (in blue)
  are obtained from numerical simulations by averaging over $10^6$
  realisations of CTRW.  The lines (in red) plot the analytical
  asymptotic results reported in the text. Upper inset plots the ratio
  of $\langle t/M \rangle$ to $l/\langle M \rangle$. Lower inset zooms
  the portion shown.}
\end{figure}

The statistics of the longest age of the records is similar to that of
the longest excursion in a renewal process that was studied
recently~~\cite{godreche:2009}. Therefore, skipping details, we cast
the result in our context. The mean longest age of records in a given
time duration $t$, for large $t$, reads $\langle l_{\max}\rangle \sim
C_\alpha\, t$, where $C_\alpha$ is given by \eref{Q}.

The mean shortest age of the records for a given $t$ is, evidently,
$\langle l_{\min}\rangle = \int_0^\infty G(u,t)\, du$, --- where
$G(u,t)=\text{Prob}[l_{\min} > u]$ is the cumulative distribution of
the shortest age. The Laplace transform of $\langle l_{\min}\rangle$
is given by
\begin{equation}
\int_0^\infty \langle l_{\min}\rangle\, e^{-st}\, dt = \int_0^\infty
du\;\frac{\int_u^\infty q(l)\, e^{-sl}\, dl}{1-\int_u^\infty f(l)\,
  e^{-sl}\, dl}, 
\end{equation}
where the integrand --- that expresses $\int_0^\infty e^{-st} G(u,t)\,
dt $ --- is obtained by taking a Laplace transform of
\eref{joint-distribution} with respect to~$t$, integrating over
$\{l_i\}$ in $(u,\infty)$ and summing over $M$.  The asymptotic
behaviour of $\langle l_{\min} \rangle$ is extracted by analysing the
leading singularity near $s=0$ in the above equation --- that yields
$\langle
l_\text{min}\rangle \sim\tau_0 \big[\Gamma(1-\alpha/2) \big]^{-1}
(t/\tau_0)^{1-\alpha/2}$ for large~$t$.

\Fref{average-fig} compares the analytical asymptotic expressions of
the averages mentioned above, with the respective values obtained from
a numerical simulation of a CTRW with $\alpha=2/3$.

In conclusion, we have shown that the statistics of records for a
correlated time series generated by a CTRW display rather rich
behaviour.  We expect the results to be relevant to a wide class of
problems. For example, CTRW has been recently used to model various
time series that occurs in finance and economics~\cite{Scalas:06}, and
hence our results would be useful in analysing those financial time
series.  For various contexts (e.g., in regard to the issues of
``climate change with a warming trend''), it may be interesting to
extend the study to non-symmetric $\phi(\xi)$ --- using the
generalised Sparre-Andersen theorem (see Ref.~\cite{Majumdar:10}).  It
is interesting to mention that for the particular case of the Cauchy
distributed jumps with a constant drift $\mu$ --- i.e., the jumps are
drawn from a non-symmetric Cauchy PDF $\phi(\xi)=\pi^{-1}
a/[a^2+(\xi-\mu)^2]$ --- one obtains similar results, namely,
$\alpha/2$ in the results of this paper is replaced by
$\alpha[1/2+\pi^{-1}\tan^{-1}(\mu/a)]$ in that case\footnote{Details
will be published elsewhere.}. Finally, it is useful to compare the
statistics of the number of records of the CTRW with that of a time
series of IID random entries drawn from $p(x)$ in continuous time with
the waiting time between successive entries drawn from
$\rho(\tau)$. In the latter case, it turns out that the distribution
of the number of records $P(M,t)$ does not depend on $p(x)$ and for
large $t$, it approaches a Gaussian around its mean $\langle
M \rangle \sim \alpha\ln(t/\tau_0)$ with a variance $\langle
M^2 \rangle - \langle M \rangle^2\sim\alpha\ln(t/\tau_0)$ ---
therefore, the qualitative features do not change drastically as one
changes $\alpha$. In contrast, for the CTRW we have found that the
qualitative behaviour depends crucially on the parameter $\alpha$.

\end{document}